\documentstyle[amssymb,aps,twocolumn,epsfig]{revtex}
\begin{document}
\draft
\title{Optoelectric spin injection in semiconductor heterostructures
without ferromagnet}
\author{A.G. Mal'shukov$^{1}$ and K.A. Chao$^{2}$}
\address{$^{1}$Institute of Spectroscopy, Russian Academy of Science,
142092 Troitsk, Moscow oblast, Russia\\
$^{2}$Solid State Theory Division, Department of Physics, Lund University,
S-223 62 Lund, Sweden}
\maketitle

\begin{abstract}
We have shown that electron spin density can be generated by a dc current
flowing across a $pn$ junction with an embedded asymmetric quantum well. Spin
polarization is created in the quantum well by radiative electron-hole
recombination when the conduction electron momentum distribution is shifted
with respect to the momentum distribution of holes in the spin split
valence subbands. Spin current appears when the spin polarization is injected
from the quantum well into the $n$-doped region of the $pn$ junction.
The accompanied emission of circularly polarized light from the quantum well
can serve as a spin polarization detector.
\end{abstract}


One of the most important problems in spintronics is the efficient injection
of spin currents into semiconductor structures. A possible way is to inject
spin polarization from a ferromagnet by passing a dc current across the
interface~\cite{aronov}. If the ferromagnet is metallic, the efficiency of
such injection is controversial~\cite{exper}, and the observed weak spin
current has been attributed to the large mismatch of spin diffusion constants
between the adjacent semiconductor and metal~\cite{wees}. Spin injection can
be enhanced by using an appropriate interface tunneling barrier~\cite{theory}.
On the other hand, rather high degree of spin current polarization has been
detected if the spin is injected from magnetic
semiconductors~\cite{fiderling}, although its high efficiency is restricted
to low temperatures.

In this Letter we propose a new method to use a dc current to inject spin
polarization, not from any ferromagnetic material, but from a quantum well
(QW) embedded into a $pn$ junction. The spin polarized current is generated
during the radiative electron-hole recombination in the QW, accompanied by the
emission of circularly polarized light. Our new injection mechanism can be
easily explained. It is well known that in a QW which is asymmetric along
the growth direction, at each finite wave vector {\bf k} the spin degeneracy
of hole subbands is removed. The corresponding splitting of hole energies
increases with $k$ and can reach a quite high value. For example, in a $p$
inversion layer of a GaAs-AlGaAs heterojunction, the splitting of the topmost
heavy-hole subband was calculated~\cite{sham} to be about 5 meV at
$k$=2$\cdot$10$^6$ cm$^{-1}$. Each of the split states is a linear combination
of four angular momentum eigenstates which are specified by the z-component
$J_z$=$\pm$3/2 and $\pm$1/2. The resulting mean spins of hole states are
parallel to the QW interfaces. For a given {\bf k} the spins in each pair of
spin-split subbands have opposite directions, and in a given spin-split
subband the mean spin at {\bf k} is equal and opposite to that at --{\bf k}.
The spin orientations in the topmost heavy-hole subband are shown in Fig.~1,
where the Fermi energy $\mu_e$ (or $\mu_h$) of the quasiequilibrium degenerate
electron gas (or hole gas) is indicated. When a $\sigma$-spin electron in the
{\bf k} state of the conduction band makes a radiative transition to the
topmost heavy-hole subband, as marked by the downward vertical arrow line in
Fig.~1, the probability of this process depends on $\sigma$ because the
population of hole at the {\bf k} state in the split (+)-subband may be
different from that in the split (--)-subband. Hence, at {\bf k} the
conduction band electron spin will be polarized along $\sigma$ direction if an
electron with (--$\sigma$)-spin has higher recombination rate. However, since
the hole spin orientation at --{\bf k} is reversed with respect to that at
{\bf k}, the spin polarization of the conduction band electron at the
--{\bf k} state will be along --$\sigma$ direction. Consequently, if the
momentum distributions of both the electron gas and the hole gas are
isotropic, there will be no net spin polarization of the conduction band
electrons.
\begin{figure}[hbt]
\begin{center}
\leavevmode
\epsfig{file=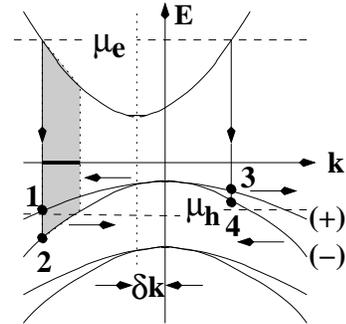, width=4.5cm}
\end{center}
\caption{Schematic illustration of the energy levels and the orientations
of mean angular momenta in the topmost heavy-hole subband in an asymmetric
quantum well.}
\label{fig1}
\end{figure}

On the other hand, if the momentum distributions of the electron gas and/or
the hole gas are anisotropic, the generation of spin polarization becomes
possible. We will explain the generating process with the help of Fig.~1,
where the anisotropic momentum distribution is indicated by a shift
$\delta${\bf k} of the quasiequilibrium momentum distribution of electrons
with respect to that of holes. The state 3 in the (+)-subband and the state
4 in the (--)-subband are equally occupied by holes with spins in opposite
directions. Hence, the total probability of recombination of a conduction
electron with the two holes in the states 3 and 4 is independent of the spin
of the electron. However, a hole appears in the state 1 in the (+)-subband but
not in the state 2 in the (--)-subband. Hence, the conduction electrons in the
shaded region can recombine only with the holes in the (+)-subband, and the
corresponding recombination probability is spin dependent. Such processes
create a nonequilibrium electron spin polarization in the QW. When this
polarization diffuses out of the QW, a spin current is then generated in the
$n$-doped region. One way to realize the situation shown in Fig.~1 is to
apply an electric field {\bf E} parallel to interfaces. Then, both electrons
and holes gain their respective drift velocities. Let {\bf v} be the relative
drift velocity of electrons with respect to drift velocity of holes. If we
ignore the drift of the low mobile holes, the resultant band positions are
illustrated in Fig.~1. The shift $\delta${\bf k} is simply
$m^{\ast}${\bf v}/$\hbar$=$m^{\ast}\mu${\bf E}/$\hbar$, where $\mu$ is the
electron mobility. The suggested device for spin current injector, a QW
embedded into a $pn$ junction, is illustrated in Fig.~2. A bias voltage $V_b$
is applied along the $z$ axis, and the transverse voltage source $V_t$ creates
an electric field {\bf E} along $x$ axis. The potential profile along $z$
axis and the spin current are also shown schematically in Fig.~2. 
\begin{figure}[hbt]
\begin{center}
\leavevmode
\epsfig{file=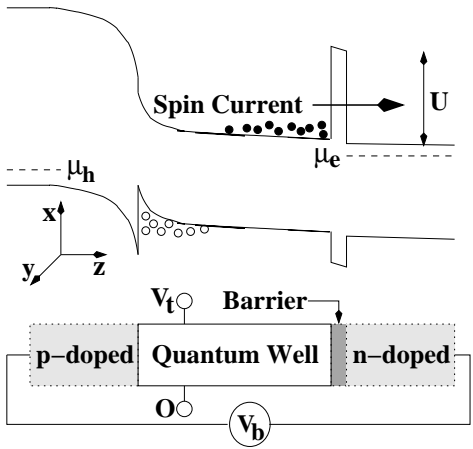, width=6.2 cm}
\end{center}
\caption{Suggested device for spin current injector. Upper plot is the
corresponding potential profile.}
\label{fig2}
\end{figure}

At low temperatures, for degenerate electron and hole gases, besides the band
to band electron transitions, excitonic recombination processes must be
considered. To avoid the complicated analysis which will not change the
essential physics, in this Letter we will neglect the excitonic effect on the
spin generation.

For the valence band we define $|J_z\rangle$ as the zone center Bloch state
and $\psi_{{\bf k,}n,J_z}^{\pm}(z)$ the $n$-th confined state in the QW
associated to the two-dimensional wave vector {\bf k} and the hole spin
projection $J_z$. Then, the wave functions $\Psi_{n,{\bf k}}^{\pm}(z)$ of the
$n$-th ($\pm$) split valence subbands can be expressed as the sum of
$\psi_{{\bf k,}n,J_z}^{\pm}(z)|J_{z}\rangle$ over $J_z$=$\pm\frac{3}{2}$ and
$\pm\frac{1}{2}$. For electrons in the lowest conduction subband, we can
similarly define $|\sigma\rangle$ and $\chi_{\bf k}(z)$. If we neglect the
small spin-orbit splitting, the electron wave functions $\Psi_{\bf k}(z)$
can be written as a linear combination of two degenerate states
$C_{\frac{1}{2}}\chi_{\bf k}(z)|\frac{1}{2}\rangle
$+$C_{-\frac{1}{2}}\chi_{\bf k}(z)|-\frac{1}{2}\rangle$, where
$C_{\pm\frac{1}{2}}$ are normalized amplitudes. When an electron and a hole
recombine from their respective quantum states to emit a photon of wave
vector {\bf q} and polarization vector {\bf e}$_{\bf q}^{\lambda}$, to the
lowest order, the quantum probability amplitude of this process can be easily
derived as
$C_{\frac{1}{2}}M_{\frac{1}{2},n}^{\lambda,\nu}({\bf k,q})$+$C_{
-\frac{1}{2}}M_{-\frac{1}{2},n}^{\lambda,\nu}({\bf k,q})$, where $\nu$=$\pm$
represents the ($\pm$) split valence subbands, and
\begin{eqnarray}\label{M}
&& M_{\sigma,n}^{\lambda,\nu}({\bf k,q}) = \nonumber \\
&& \frac{eA_q}{mc} \sum_{J_z}
{\bf p}_{\sigma,J_z}\cdot {\bf e}_{\bf q}^{\lambda}
\int dz \,\psi_{{\bf k},n,J_z}^{\nu\ast}(z)\chi_{\bf k}(z) \, .
\end{eqnarray}
In the above equation,
{\bf p}$_{\sigma,J_z}$$\equiv$$\langle J_z|{\bf p}|\sigma\rangle$ is the
interband matrix element of the electron momentum operator {\bf p}, and
$A_q$$\equiv$$\sqrt{2\pi c/nq}$.

Knowing the recombination probability of the conduction band electrons with a
specific spin, the spin generation during the recombination process can be
calculated. The resultant spin density in the conduction band is derived as
\begin{equation}\label{S}
{\bf G}_{s} = - \sum_{\sigma\sigma^{\prime}}
G_{\sigma\sigma^{\prime}}{\bf s}_{\sigma^{\prime}\sigma} \, ,
\end{equation}
where 2{\bf s}$_{\sigma^{\prime}\sigma}$ are Pauli matrices. The matrix
elements $G_{\sigma\sigma^{\prime}}$ have the form
\begin{equation}\label{G}
G_{\sigma\sigma^{\prime}} = A \sum_{{\bf k,q},\lambda,n,\nu}
M_{\sigma^{\prime},n}^{\lambda,\nu\ast}({\bf k,q})
M_{\sigma,n}^{\lambda,\nu}({\bf k,q}) f({\bf k}) f_n^{\nu}({\bf k}) \, ,
\end{equation}
where $A$ is a numerical factor, and $f({\bf k})$ [$f_n^{\nu}({\bf k})$] is
the momentum distribution functions of electrons (or holes). Within the
perturbation theory, $f({\bf k})$ and $f_n^{\nu}({\bf k})$ are assumed to be
spin independent. A relevant dimensionless parameter which measures the
efficiency of spin generation is {\bf P}={\bf G}$_s/G$, where
$G$$\equiv$$G_{\frac{1}{2},\frac{1}{2}}$+$G_{-\frac{1}{2},-\frac{1}{2}}$ is
the number of radiative recombinations per the time unit .

As mentioned earlier, we will consider higher temperature such that the
excitonic effect can be neglected. Hence, each momentum distribution function
is the sum of the equilibrium Boltzmann distribution and the nonequilibrium
correction due to the electric field {\bf E}. Since the drift of holes can be
ignored, we have
\begin{eqnarray}\label{n}
f_n^{\nu}({\bf k}) &=&
\exp{ \left[ \frac{\mu_h-\epsilon_n^{\nu}(k)}{k_BT}\right] } \, , \nonumber \\
f({\bf k}) &=&
\exp{ \left[ \frac{\mu_e-\epsilon(k)}{k_BT}\right] }
\left( 1-\frac{\hbar{\bf v}\cdot {\bf k}}{k_BT}\right) \, ,
\end{eqnarray}
where $\epsilon(k)$ is the lowest conduction subband, and $\epsilon_n^{\nu}(k)$
$(\pm)$ split valence subbands. The zero reference energy is set at the
bottom of the lowest conduction subband.

We need the valence subband wave functions $\psi_{{\bf k},n,J_z}^{\pm}(z)$
for calculating of $M_{\sigma,n}^{\lambda,\nu}({\bf k})$. Let the growth
direction $z$ be along the [001] crystal. The wave functions can be obtained
by applying a proper unitary transformation\cite{sham} which block
diagonalizes the Luttinger Hamiltonian. They can be expressed in the general
form as
\begin{eqnarray}\label{vector}
\psi_{{\bf k},n,-\frac{3}{2}}^{\pm \ast}(z) &=&
\pm \psi_{{\bf k},n,\frac{3}{2}}^{\pm}(z) =
\pm\, i \,\exp{(i\phi_{\bf k})}\, \xi_{nh}^{\pm}(z)/\sqrt{2} \, , \nonumber \\
\psi_{{\bf k},n,-\frac{1}{2}}^{\pm \ast}(z) &=&
\pm \psi_{{\bf k},n,\frac{1}{2}}^{\pm}(z) =
\pm\, i \,\exp{(i\eta_{\bf k})}\, \xi_{nl}^{\pm}(z)/\sqrt{2} \, ,
\end{eqnarray}
where the real functions $\xi_{nh}^{\pm}(z)$ and $\xi_{nl}^{\pm}(z)$ represent
the partial amplitudes of heavy and light holes in the (+)- and the
(--)-state. The phases are
$\phi_{\bf k}$=3$\eta_{\bf k}$=--3$\varphi_{\bf k}/2$ with
$\varphi_{\bf k}$=$\cos^{-1}(k_x/k)$.
To obtain (\ref{vector}) we have neglected the band warping. For convenience
we set the x-axis along the drift velocity direction.

With all above equations (\ref{M})-(\ref{vector}) we are ready to calculate
the efficiency of spin generation {\bf P}={\bf G}$_s/G$. In terms of the
overlap integrals $b_{nh}^{\pm}$=$\int dz\xi_{nh}^{\pm}(z)\chi_{\bf k}(z)$
and $b_{nl}^{\pm}$=$\int dz\xi_{nl}^{\pm}(z)\chi_{\bf k}(z)$, we define
$R_{x,nk}^{\pm}$=$(b_{nl}^{\pm})^2/3$,
$R_{y,nk}^{\pm}$=--$b_{nl}^{\pm}b_{nh}^{\pm}/\sqrt{3}$, and
$R_{z,nk}^{\pm}$=0. Then, {\bf P} is derived as
\begin{eqnarray}\label{P}
{\bf P} &=& \frac{\hbar v}{2k_BT}
\left( \sum_{{\bf k},n,\nu=\pm} \nu k \, {\bf R}_{nk}^{\nu} \,
{\cal F}_{k,n,\nu} \right) \nonumber \\
&& \times \left( \sum_{{\bf k},n,\nu=\pm}
[(b_{nh}^{\nu})^2 + (b_{nl}^{\nu})^2/3] \,
{\cal F}_{k,n,\nu} \right)^{-1} \, ,
\end{eqnarray}
where ${\cal F}_{k,n,\nu}$=$\exp{\{-[\epsilon_n^{\nu}(k)+\epsilon(k)]/
k_BT\}}$. As expected, only the anisotropic part ({\bf v}$\cdot${\bf k} term)
of the electron momentum distribution function $f({\bf k})$ has contributed to
the generated polarization.

Based on the above formula, {\bf P} can be investigated numerically in a broad
range of parameters. However, qualitative results can be derived from
(\ref{P}) analytically for $kd$$\ll$1, where $d$ is the typical length of hole
confinement in the QW.  We notice that in (\ref{P}), at a temperature $T$, a
characteristic wave vector $k_T$ can be defined as the mean value of $k$ with
respect to the Boltzmann factor ${\cal F}_{k,n,\nu}$. Then, an order of
magnitude evaluation of {\bf P} can be obtained at $k_{T}d$$\ll$1. Our
calculation indicates that in the region $kd$$\ll$1, the energy split
$\Delta_n(k)$=$\epsilon_n^+(k)$-$\epsilon_n^-(k)$ is much less than the
separation of two adjacent hole subbands, and $b_{nl}^+$$\simeq$$b_{nl}^-$ as
well as $b_{nh}^+$$\simeq$$b_{nh}^-$. Hence,
${\bf R}_{nk}^+$$\simeq$${\bf R}_{nk}^-$, and the term
(${\bf R}_{nk}^+{\cal F}_{k,n,+}$--${\bf R}_{nk}^-{\cal F}_{k,n,-}$) to be
summed up in the numerator of (\ref{P}) becomes proportional to
1--$\exp{[-\Delta_n(k)/k_BT]}$. With such simplification we can analyze the
physical processes which contribute to {\bf P}. Let us consider the
contribution of the lowest hole subband with $n$=1. If this is a heavy hole
subband, then the small admixture of light hole states to this subband allows
us to evaluate $b_{1l}$ to obtain $b_{1l}$$\propto$$(kd)^2b_{1h}$. As a
result, we have $R_{y,1k}$$\gg$$R_{x,1k}$, and the generated spin
polarization is oriented perpendicular to the drift direction of electrons.
On the other hand, if the $n$=1 subband corresponds to the light hole, 
similar analysis gives $b_{1h}$$\propto$$(kd)^2b_{1l}$ and
$R_{x,1k}$$\gg$$R_{y,1k}$. Consequently, the so generated spin polarization is
oriented parallel to the electron drift direction.

In the range of temperatures where $\Delta_1(k_T)/k_BT$$\ll$1, we derive
from (\ref{P}) 
\begin{equation}\label{Py}
{\cal P} \equiv |{\bf P}| \propto \frac{\hbar vk_T}{k_BT} \cdot
\frac{\Delta_1(k_T)}{k_BT} \, (k_Td)^{\beta}\ ,
\end{equation}
where $\beta$=2 (or $\beta$=0) if the lowest subband is of heavy (or light)
hole type. Hence, in the region of small characteristic wave vectors, the
optoelectric generation of spin polarization is more efficient if the lowest
hole subband is of the light hole type. We should mention that in III-V
semiconductor QW, the lowest hole subband is of light hole type if the QW is
sufficiently strained. One example is InAs/GaAs QW.

In (\ref{Py}) the factor $\hbar vk_T/k_BT$ is due to the anisotropic momentum
distribution of the electrons. The other factor
$\Delta_1(k_T)(k_Td)^{\beta}/k_BT$ is originated from the ratio of the two
summations in (\ref{P}), and increases with $T$ through $k_T$ and
$\Delta_1(k_T)$. For example, in GaAs at room temperature,
$k_T$$\simeq$$2\cdot 10^6$~cm$^{-1}$, and then $k_Td$$\simeq$1 for
$d$=100~{\AA}. Under this situation, $\Delta _1(k_T)$ becomes comparable to
the quantization energy of hole subbands~\cite{sham}, and (\ref{Py}) is no 
longer valid. However, a simple scaling analysis shows that around
$k_Td$$\simeq$1, Eq.(\ref{Py}) can still be used to evaluate ${\cal P}$ with
the factor $(k_Td)^{\beta}$ replaced by a numerical factor of the order of
unity. 

We have shown that spin polarization of conduction electrons can be generated
in a quantum well due to the radiative electron-hole recombination. Let
$|e|I$ be the electric current across the $pn$ junction, and
$\eta$=$\tau_{nr}/(\tau_{nr}+\tau_r)$ the luminescence quantum efficiency,
where $\tau_r$ (or $\tau_{nr}$) is the radiative (or nonradiative)
electron-hole recombination time. Then, the number of radiative
recombinations per time unit is $G$=$I\eta$. The so generated spin
polarization in the conduction band can either diffuse out of the QW into the
$n$-doped region of the $pn$ junction, or relax within the QW with a spin
relaxation time $\tau_{sw}$. In III-V semiconductor QWs the dominating
process is the the D'yakonov-Perel' spin relaxation~\cite{bir}, although the
Bir-Pikus mechanism~\cite{bir} can be efficient due to presence of a large
number of holes. In the steady state, the spin current which flows out of the
QW is given as
\begin{equation}\label{is}
I_s = {\cal P}I\eta - S/\tau - S/\tau_{sw} \, ,
\end{equation}
where $S$ is the spin polarization, and 1/$\tau$=1/$\tau_r$+1/$\tau_{nr}$.
We choose the spin quantization axis parallel to the direction of the vector
{\bf P} defined in (\ref{P}). Then, $S$ is simply
($n_{\frac{1}{2}}$-$n_{-\frac{1}{2}}$)/2, where $n_{\sigma}$ is sheet
electron density in the QW with $\sigma$-spin.

To relate $S$ to $I$, we need to specify the transport process between the QW
and the $n$-doped region, which involves the spin diffusion and relaxation in
the $n$-doped semiconductor. To avoid such complication, we will consider a
simple model of thermionic transport over the barrier, which is suitable for
not too low temperatures. In this case the $\sigma$-spin current $i_{\sigma}$
is determined by the balance of Richardson currents emitted from both sides
of the potential barrier. Such emission currents depend on the chemical
potentials $\mu_{\sigma w}$ in the QW and $\mu_{\sigma b}$ in the $n$-doped
bulk semiconductor. In the linear response regime,
$|\mu_{\sigma w}$--$\mu_{\sigma b}|$$\ll$$k_BT$, and we have
\begin{equation}\label{irich}
i_{\sigma} = \frac{A^{\ast}T^2}{|e|} \cdot
\frac{\mu_{\sigma w}-\mu_{\sigma b}}{k_BT} \,
\exp{(\frac{\mu_{\sigma b}-eU}{k_BT})} \, ,
\end{equation}
where $A^{\ast}$ is the Richardson constant, and the the barrier height $U$
is indicated in Fig.~2. With this expression, the spin current
$I_s$=($i_{\frac{1}{2}}$--$i_{-\frac{1}{2}}$)/2 is a function of the
chemical potentials.

In the $n$-doped bulk semiconductor we assume diffusive motion of electrons
with a diffusion constant $D$ and a spin relaxation time $\tau_{sb}$. We
should mention that within the framework of linear transport theory the spin
current in a bulk material can be driven only by a spin density gradient, and
the corresponding characteristic length of the spin density variation has the
form $L_s$=$\sqrt{D\tau_{sb}}$. In an n-doped bulk semiconductors $\tau_{sb}$
can be very long~\cite{kikkawa}, and a large $L_s$ will reduce the efficiency
of spin injection. This is the same problem encountered in the study of spin
injection from ferromagnetic metals into semiconductors~\cite{wees}, which
can be overcome~\cite{theory} with a proper choice of the barrier height $U$.
If $U$ is so chosen that ($L_s/l)\exp{[(\mu_{\sigma b}-eU)/k_BT]}$$\ll$1,
where $l$ is the electron mean free path, the injected spin current is no
longer limited by the low spin relaxation rate of the bulk semiconductor. For
a moderately doped III-V semiconductors, if we take
from~\onlinecite{bir,kikkawa} $\tau_{sb}$=100~ps and the electron mobility
5$\cdot$10$^4$cm$^2$/Vsec, $L_s/l$$\simeq$10 at room temperature. Therefore,
the above inequality can be easily satisfied even if the barrier height is
rather low. In this situation we obtain
\begin{equation}\label{is2}
I_s = I{\cal P}\eta \left[  1 + \frac{\Delta\mu}{k_BT}
(1+\frac{\tau}{\tau_{sw}}) \right] ^{-1} \, ,
\end{equation}%
where $\Delta\mu$=($\mu_{\frac{1}{2},w}$+$\mu_{-\frac{1}{2},w}$--$\mu_{
\frac{1}{2},b}$--$\mu_{-\frac{1}{2},b}$)/2.

Although $\Delta\mu/k_BT$ is assumed to be small in linear response theory,
the ratio $\tau/\tau_{sw}$ can be large, and so the second term in the square
brackets of (\ref{is2}) can not be neglected. For example, in bulk III-V
semiconductors with carrier density in the range between 10$^{17}$~cm$^{-3}$
and 10$^{18}$~cm$^{-3}$, at room temperatures the ratio $\tau/\tau_{sw}$ is
about ten~\cite{bir,kikkawa}, assuming $\tau$ in the range of nanosecond. The
similar ratio was also found in QWs~\cite{sham2}. In our system, as
illustrated in Fig~2, the electron density and the hole density in the
asymmetric QW are spatially shifted with respect to each other. The
electron-hole recombination becomes spatially indirect, and therefore $\tau$
increases. On the other hand, $\tau_{sw}$ increases if the spin relaxation is
dominated by the Bir-Pikus mechanism, but remains almost the same if the
D'yakonov-Perel' mechanism dominates. Consequently, the actual value of
$\tau/\tau_{sw}$ varies with the system to be studied. From (\ref{is2}), we
see that the upper limit of the spin polarization of the injected current is
$I_{s}/I$=${\cal P}\eta$.

Since the spin polarization in the QW is created by the emission of
circularly polarized light, the spin generation can be detected by measuring
the polarization of emitted photons. Circular polarization of a photon is
represented by the imaginary off-diagonal elements of the photon polarization
matrix $\rho$. This Hermitian matrix can be calculated in a way similar to
the calculation of spin polarization. It can be shown that the imaginary
elements are related to {\bf P} as $\rho_{xz}$$\propto$$iP_y$ and
$\rho_{yz}$$\propto$--$iP_x$. Hence, most of the circularly polarized photons
are emitted with their wave vectors parallel to {\bf P}.

If we introduce a ferromagnetic layer to the $n$-doped part of our sample,
the optoelectric injection of spins can be investigated by measuring the
resistance of the system. The separation of the ferromagnetic layer and the
QW must be less than $L_s$. In this case, the measured resistance depends on
relative spin polarizations in the paramagnetic and the ferromagnetic
materials~\cite{aronov,exper}, and can be varied by changing the direction
of the applied electric field {\bf E}.

We close this Letter with one remark. Within the framework of linear response
theory used in our analysis, the ratio $\hbar vk_T/k_BT$ must be small. With
a stronger electric field to increase the electron drift velocity $v$, we
conjecture an enhancement of the spin generation under an anisotropic
momentum distribution of hot electrons.

A. G. M. acknowledges the support from the Royal Swedish Academy of Sciences,
and from the Crafoord Foundation.

\end{document}